\documentclass[pra,superscriptaddress]{revtex4}
\usepackage{amssymb}
\usepackage{amsmath}
\usepackage{graphicx}
\usepackage{epstopdf}
\usepackage{hyperref}

\newtheorem{theorem}{Theorem}

\newtheorem{corollary}{Corollary}

\def\be{\begin{eqnarray}}
\def\ee{\end{eqnarray}}
\def\Tr{\mathrm{Tr}}
\def\I{\mathbb{I}}
\begin{document}

\title{Detecting EPR steering via two classes of local measurements}
\author{Le-Min Lai}
%\thanks{lailemin@outlook.com.}
\affiliation{School of Mathematical Sciences,  Capital Normal University,  Beijing 100048,  China}
\author{Tao Li}
\thanks{litao@btbu.edu.cn.}
\affiliation{School of Mathematics and Statistics, Beijing Technology and Business University, Beijing 100048, China}
\author{Shao-Ming Fei}
\thanks{feishm@cnu.edu.cn}
\affiliation{School of Mathematical Sciences,  Capital Normal University,  Beijing 100048,  China}
\affiliation{Max Planck Institute for Mathematics in the Sciences, 04103, Leipzig, Germany}
\author{Zhi-Xi Wang}
\thanks{wangzhx@cnu.edu.cn}
\affiliation{School of Mathematical Sciences,   Capital Normal University,   Beijing 100048,   China}

\begin{abstract}
\baselineskip18pt
We study the Einstein-Podolsky-Rosen (EPR) steering and present steerability criteria for arbitrary qubit-qudit (qudit-qubit) systems based on mutually unbiased measurements (MUMs) and general symmetric informationally complete measurements (general SIC-POVMs). Avoiding the usual complicated steering inequalities, these criteria can be more operational than some existing criteria and  implemented experimentally. Detailed examples are given to illustrate the efficiency of the criteria in both computation and experimental implementation.
\end{abstract}

\maketitle

\section{Introduction}

As a distinctive and key feature in quantum world, the nonlocality challenges our intuition and comprehension about the nature. In the heart of nonlocality is the concept ``EPR paradox'' raised by Einstein, Podolsky, and Rosen in their seminal paper \cite{epr}, which indicates that there were some conflicts between quantum mechanics and local realism. They proposed the possible existence
of ``local hidden variable'' (LHV) models. With respect to the EPR paradox, Schr\"{o}dinger introduced the concept ``steering'' \cite{sch1, sch2} to characterize the Alice's ability of remotely steering Bob's state by local measurements. These counterintuitive nonlocal effects, or ``spooky action at a distance,''  were collectively dubbed ``entanglement''. In 1964, Bell introduced his famous inequality for local hidden variable theories, which crucially brought the nonlocality debate to an experimentally testable form  \cite{Bell}. Thus, three distinct types of nonlocal correlations: entanglement, Schr\"{o}dinger¡¯s steering and Bell nonlocality, were intuitively elaborated, which have opened an epoch of unrelenting exploration of quantum correlations.

The nonlocality and quantum entanglement play important roles in our
fundamental understandings of physical world as well as in various
novel quantum informational tasks \cite{nielsen}.
A bipartite quantum state admits no LHV models if it violates some Bell inequalities such that the local measurement
outcomes can not be modeled by classical random distributions over probability spaces, termed as Bell nonlocal \cite{chsh, bcp}.
A quantum state without entanglement must admit LHV models. However, not all
the entangled quantum states are of nonlocality \cite{we,rpm}.

Entanglement and Bell nonlocality have attained flourishing developments.
The concept of EPR steering was only  introduced in 2007 in the form of a quantum task \cite{WJD}.
The task of quantum steering is that a referee has to determine, by using the measurement outcomes communicated classically from the two parties to the referee,
whether two spatially separated parties share entanglement, when one of the two parties is untrusted.
The notion of EPR steering was introduced as the inability to construct a local hidden state (LHS) model to explain the joint probabilities of measurement outcomes.
It has been shown that EPR steering is an intermediate between entanglement and Bell nonlocality.
According to the hierarchy of nonlocality, the set of steerable states is a strict subset of entangled states and a strict superset of Bell nonlocal states  \cite{bells}.
Moreover, the EPR steering is inherently asymmetric with respect to the observers, unlike quantum nonlocality and entanglement \cite{one1}.
There exist entangled states which are one-way steerable, demonstrating steerability from one observer to another spatially separated observer, but not vice-versa \cite{one1, one2, one3}.

EPR steering not only has foundational significance of describing the nonlocality, but also has a vast range of information theoretic applications, ranging from one-sided device-independent quantum key distribution \cite{app1}, advantages in sub-channel discrimination \cite{app2}, secure quantum teleportation \cite{app3}, quantum communication \cite{app4}, detecting bound entanglement \cite{app5}, one-sided device-independent randomness generation \cite{app6}, to one-sided device-independent self-testing of pure maximally as well as non-maximally entangled states \cite{app7}.

Against the above backdrop, from a fundamental viewpoint as well as an information-theoretic perspective, it is important to detect whether a quantum state is steerable or not. A number of EPR steering criteria  have been proposed till date \cite{in1, in2, in3, in4, in5, in6, in7, in8, in9, in10, in11, in12, in13, in14, in15}. Recently, in \cite{steer1, steer2}, the authors focused on detecting arbitrary qubit-qudit state $\rho_{AB}$ and gave a criterion by detecting the entanglement of a new constructed state, $\mu\rho_{AB}+(1-\mu)\frac{\I}{2}\otimes \rho_{B}$, without using any steering inequality. Following the positive partial transposition criterion \cite{ph1, ph2}, the authors in \cite{steer2} present a brief idea on how their result can be implemented in experiments for two-qubit states. Although such result consumes some resources, it provides a way of detecting EPR steering by avoiding steering inequalities.

In \cite{mum} the authors formulated an effective tool called mutually unbiased measurements (MUMs) to study the problem of quantum entanglement. Besides, there is another useful tool called general symmetric informationally complete measurements (general SIC-POVMs) \cite{gsic1, gsic2}. Both the MUMs and general SIC-POVMs can be used to detect quantum entanglement \cite{crti15, crti16, crti17, crti18, crti19, crti20}. These entanglement criteria are shown to be powerful and can be implemented experimentally. Due to the relationship between the entanglement and the EPR steering, we present steering criteria in terms of MUMs and general SIC-POVMs.

\section{Detection of EPR Steering}\label{sec2}

The EPR steering is usually formulated by considering a quantum information task \cite{bells,WJD}. Suppose two spatially separated observers, say Alice
and Bob, want to share entanglement between each other. Alice prepares a bipartite quantum state $\rho_{AB}$ and sends one partite to Bob.
Bob trusts his own but not Alice's apparatus. He will be convinced that they share an entangled state only if there exists evidence that Alice can ``steer'' Bob's state
by performing measurements on their respective subsystems. If Alice (Bob) performs a measurement $A$ ($B$) with measurement outcomes $a$ ($b$) on her (his) system,
the joint probability of obtaining the outcomes $a$ and $b$ is given by
\be
P(a, b|A, B; \rho_{AB})=\Tr[(\mathrm{\Pi}_a^A\otimes\mathrm{\Pi}_b^B)\rho_{AB}],
\ee
where $\mathrm{\Pi}_a^A$ and $\mathrm{\Pi}_b^B$ are the corresponding projective operators for Alice and Bob, respectively.

The only way that the dishonest Alice pretends to steer Bob¡¯s state, is to send some local hidden states (LHS) with ensemble $\{p_\lambda \rho_\lambda\}$,
where $\lambda$ is the hidden variable, $\rho_\lambda$ is the state that Alice sends with probability $p_\lambda$ $(\sum_{\lambda} p_\lambda = 1$).
She announces an outcome according to her knowledge about the sent states. In this case the correlation will be of the form
\begin{equation}\label{lhv-lhs}
P(a, b|A, B; \rho_{AB}) = \sum_{\lambda} p_\lambda \, P(a|A,\lambda) \, P_Q(b|B,\rho_{\lambda}),
\end{equation}
where $P(a|A,\lambda)$ can be any possible probability distribution
that Alice designed, $P_Q(b|B,\rho_{\lambda})=\Tr [\mathrm{\Pi}_b^B\rho_{\lambda}]$ denotes the quantum probability of outcome $b$ given by measuring $B$ on the local hidden state $\rho_{\lambda}$.
If Bob finds that any LHS models fail to satisfy such correlation Eq. (\ref{lhv-lhs}), he has to admit that Alice can steer his system and
the corresponding bipartite state is entangled.
In short, the bipartite state  $\rho_{AB}$ is unsteerable by Alice to Bob if and only if the joint probability distributions satisfy
the relation (\ref{lhv-lhs}) for all measurements $A$ and $B$.

\subsection{Detecting EPR Steering via Mutually Unbiased Measurements}

Two orthonormal bases $\mathcal{B}_1=\{|i\rangle\}^{d}_{i=1}$ and $\mathcal{B}_2=\{|j\rangle\}^{d}_{j=1}$ of $\mathbb{C}^d$ are said to be mutually unbiased if
\be
|\langle i|j\rangle|=\frac{1}{\sqrt{d}},~~~~~~\mathrm{for~~all}~~ i,j=1,2,\cdots,d.
\ee
A set of orthonormal bases $\{\mathcal{B}_1,\mathcal{B}_2,\cdots,\mathcal{B}_m\}$ in $\mathbb{C}^d$ is called a set of mutually unbiased bases (MUBs) \cite{mub} if every pair of bases in the set is mutually unbiased. In a  $d$ dimensional Hilbert space, there are at most $d+1$ pairwise unbiased bases. This set is called a complete set of MUBs. It is still an open problem whether complete set of MUBs exists for arbitrary $d$.

In Ref. \cite{mum}, the authors introduced the concept of MUMs. Two POVM measurements on $\mathbb{C}^d$ , $\mathcal{P}^{(b)}=\{P_n^{(b)}\}_{n=1}^{d}$, $b=1,2$, are said to be mutually unbiased measurements if
\be
&&\Tr[P_n^{(b)}]=1,\nonumber\\[2mm]
&&\Tr[P_n^{(b)}P_{n^{'}}^{(b^{'})}]=\frac{1}{d},~~~   b\neq b^{'}\nonumber\\
&&\Tr[P_n^{(b)}P_{n^{'}}^{(b)}]=\delta_{n{n^{'}}}\kappa+(1-\delta_{n{n^{'}}})\frac{1-\kappa}{d-1},
\ee
where $1/d< \kappa \leqslant 1$, and $\kappa=1$ if and only if $\mathcal{P}^{(1)}$ and  $\mathcal{P}^{(2)}$ reduce to projective measurements with respect to two MUBs.

A general construction of $d+1$ MUMs has been presented
in \cite{mum}. Let $\{F_{n,b}: n =1,2,\cdots, d-1,  b=1 ,2,\cdots, d+1\}$ be a set of $d^2-1$ Hermitian and traceless operators acting on $\mathbb{C}^d$, satisfying $\Tr(F_{n,b}F_{n^{\prime},b^{\prime}})=\delta_{nn^{\prime}}\delta_{bb^{\prime}}$. Define $d(d+1)$ operators
\be
F_{n}^{(b)}=\left\{
              \begin{array}{ll}
                F^{(b)}-(d+\sqrt{d})F_{n,b}, & n=1,2,\cdots, d-1, \\[2mm]
                (1+\sqrt{d})F^{(b)}, & n=d,
              \end{array}
            \right.
\ee
where $F^{(b)}=\sum\limits^{d-1}\limits_{n=1}F_{n,b}$, $b=1,2,\cdots, d+1$. Then the $d + 1$ MUMs are given by
\be \label{MUMs}
P^{(b)}_{n}=\frac{1}{d}\I+tF_{n}^{(b)},
\ee
with $b=1,2,\cdots, d+1$, $n =1,2,\cdots, d$, and $t$ is so chosen such that $P^{(b)}_{n}\geqslant0$.
$d + 1$ MUMs can be expressed in such form for any dimension $d$.

Now we study the steering criteria for qudit-qubit and qubit-qudit quantum systems based on MUMs.
\begin{theorem}\label{thm:1}
Let $\rho_{AB}$ be a qudit-qubit state in $\mathbb{C}^{d}\otimes\mathbb{C}^{2}$ shared by Alice and Bob, and $\{\mathcal{P}^{(b)}\}_{b=1}^{d+1}$ and $\{\mathcal{Q}^{(b)}\}_{b=1}^{3}$ be any two  complete MUMs on $\mathbb{C}^{d}$ and $\mathbb{C}^{2}$ with the parameter $\kappa_1$ and $\kappa_2$, respectively, where $\mathcal{P}^{(b)}=\{P_n^{(b)}\}_{n=1}^d$ and $\mathcal{Q}^{(b)}=\{Q_n^{(b)}\}_{n=1}^2$. Set
$R_n^{(b)}=Q_n^{(b)}$ for $1\leqslant b \leqslant 3$, $1\leqslant n \leqslant 2$, and
$R_n^{(b)}= \frac{\I}{2}$ for $3< b \leqslant d+1$ or $3\leqslant n\leqslant d$.
Define $J(\rho)=\sum\limits_{b=1}\limits^{d+1}\sum\limits_{n=1}\limits^{d} \Tr[(P_n^{(b)}\otimes R_n^{(b)})\rho]$. If
\be\label{thm1}
J(\rho_{AB})> \frac{\sqrt{\kappa_1+1}\sqrt{4\kappa_2+4+(d+3)(d-2)}}{2\mu}-\frac{(d+1)(1-\mu)}{2\mu},
\ee
then $\rho_{AB}$ is steerable from Bob to Alice, where $\mu\in (0,\frac{1}{\sqrt{3}}]$.
\end{theorem}

\textit{Proof.} Denote $\tau_{AB}=\mu\rho_{AB}+(1-\mu)\rho_A\otimes\frac{\I}{2}$, where $\rho_A=\Tr_B[\rho_{AB}]$ is the reduced state at Alice's side. we have
\begin{eqnarray*}
\small
 J(\tau_{AB})&=&\sum\limits_{b=1}\limits^{d+1}\sum\limits_{n=1}\limits^{d} \Tr[(P_n^{(b)}\otimes R_n^{(b)})\tau_{AB}]\\
&=&\sum\limits_{b=1}\limits^{d+1}\sum\limits_{n=1}\limits^{d} \Tr[(P_n^{(b)}\otimes R_n^{(b)})(\mu\rho_{AB}+(1-\mu)\rho_A\otimes\frac{\I}{2})]\\
&=&\mu\sum\limits_{b=1}\limits^{d+1}\sum\limits_{n=1}\limits^{d} \Tr[(P_n^{(b)}\otimes R_n^{(b)})(\rho_{AB})]+(1-\mu)\sum\limits_{b=1}\limits^{d+1}\sum\limits_{n=1}\limits^{d} \Tr[(P_n^{(b)}\otimes R_n^{(b)})(\rho_A\otimes\frac{\I}{2})]\\
&=&\mu\sum\limits_{b=1}\limits^{d+1}\sum\limits_{n=1}\limits^{d} \Tr[(P_n^{(b)}\otimes R_n^{(b)})(\rho_{AB})]+(1-\mu)\sum\limits_{b=1}\limits^{d+1}\sum\limits_{n=1}\limits^{d} \Tr[P_n^{(b)}\rho_A\otimes R_n^{(b)}\frac{\I}{2}]\\
&=&\mu\sum\limits_{b=1}\limits^{d+1}\sum\limits_{n=1}\limits^{d} \Tr[(P_n^{(b)}\otimes R_n^{(b)})(\rho_{AB})]+(1-\mu)\sum\limits_{b=1}\limits^{d+1}\sum\limits_{n=1}\limits^{d} \Tr[P_n^{(b)}\rho_A]\Tr[R_n^{(b)}\frac{\I}{2}]\\
&=&
 \mu J(\rho_{AB})+(1-\mu)\sum\limits_{b=1}\limits^{3}\{\sum\limits_{n=1}\limits^{2} \Tr[P_n^{(b)}\rho_A]\Tr[Q_n^{(b)}\frac{\I}{2}]+\sum\limits_{n=3}\limits^{d} \Tr[P_n^{(b)}\rho_A]\Tr[\frac{\I^2}{4}]\}\\
 &~~~~~~+&(1-\mu)\sum\limits_{b=4}\limits^{d+1}\sum\limits_{n=1}\limits^{d} \Tr[P_n^{(b)}\rho_A]\Tr[\frac{\I^2}{4}]\\
&=&\mu J(\rho_{AB})+\frac{(d+1)(1-\mu)}{2}\\
&>& \sqrt{\kappa_1+1}\sqrt{\kappa_2+1+\frac{(d+3)(d-2)}{4}}.
\end{eqnarray*}
The last inequality follows from (\ref{thm1}).

In \cite{crti20}, the authors presented a separability criterion: if a bipartite state $\tau_{AB}$ in $\mathbb{C}^{d_1}\otimes\mathbb{C}^{d_2}$ $(d_1\geqslant d_2)$ is separable, one has $J(\tau)\leqslant\sqrt{\kappa_1+1}\sqrt{\kappa_2+1+\frac{(d_1-d_2)(d_1+d_2+1)}{4}}$. In particular, for the case $d_1=d$ and $d_2=2$, one has
$J(\tau)\leqslant\sqrt{\kappa_1+1}\sqrt{\kappa_2+1+\frac{(d+3)(d-2)}{4}}$ for all separable states $\tau$ in $\mathbb{C}^{d}\otimes\mathbb{C}^{2}$.
From this criterion, we have that $\tau_{AB}$ must be entangled.
In addition, from that any qudit-qubit state $\rho_{AB}$ is EPR steering from Bob to Alice if the state $\tau_{AB}=\mu\rho_{AB}+(1-\mu)\rho_A\otimes\frac{\I}{2}$ is entangled \cite{steer1, steer2}, we complete the proof. $\blacksquare$

On the other hand, for a qubit-qudit state $\rho_{AB}$  in $\mathbb{C}^{2}\otimes\mathbb{C}^{d}$ shared by Alice and Bob, we can detect the EPR steering from Alice to Bob through the following theorem.

\begin{theorem}\label{thm:2} Let
$\{\mathcal{P}^{(b)}\}_{b=1}^{3}$  and $\{\mathcal{Q}^{(b)}\}_{b=1}^{d+1}$ be  any two  complete MUMs on $\mathbb{C}^{2}$ and $\mathbb{C}^{d}$ with the  parameter $\kappa_1$ and $\kappa_2$, respectively, where $\mathcal{P}^{(b)}=\{P_n^{(b)}\}_{n=1}^2$, $\mathcal{Q}^{(b)}=\{Q_n^{(b)}\}_{n=1}^d$. Set $R_n^{(b)}=P_n^{(b)}$, for $1\leqslant b \leqslant 3$, $1\leqslant n \leqslant 2$, and $R_n^{(b)}=\frac{\I}{2}$ for $3< b \leqslant d+1$ or $3\leqslant n \leqslant d$.
For a qubit-qudit state $\rho_{AB}$ in $\mathbb{C}^{2}\otimes\mathbb{C}^{d}$ shared by Alice and Bob, if
\be
J(\rho_{AB})=\sum\limits_{b=1}\limits^{d+1}\sum\limits_{n=1}\limits^{d} \Tr[(R_n^{(b)}\otimes Q_n^{(b)})\rho_{AB}]>
 \frac{\sqrt{4\kappa_1+4+(d+3)(d-2)}\sqrt{\kappa_2+1}}{2\mu}-\frac{(d+1)(1-\mu)}{2\mu},
\ee
then $\rho_{AB}$ is steerable from Alice to Bob,  where $\mu\in (0,\frac{1}{\sqrt{3}}]$.
\end{theorem}

The proof is similar to that of Theorem \ref{thm:1}, by defining the state $\sigma_{AB}=\mu\rho_{AB}+(1-\mu)\frac{\I}{2}\otimes\rho_B$, where $\rho_B=\Tr_A[\rho_{AB}]$ is the reduced state at Bob's side.

As a particular case, let us consider a two-qubit state $\rho_{AB}$ in $\mathbb{C}^{2}\otimes\mathbb{C}^{2}$. Denote $\sigma_1$, $\sigma_2$ and $\sigma_3$ the Pauli matrices. We have the following corollary:

\begin{corollary} \label{cor:1}
Set $H(\rho_{AB})=\Tr[(\sigma_1\otimes\sigma_1-\sigma_2\otimes\sigma_2+\sigma_3\otimes\sigma_3)\rho_{AB}]$. If $H(\rho_{AB})>\displaystyle\frac{1}{\mu}$, $\mu\in (0,\frac{1}{\sqrt{3}}]$, then the qubit-qubit state
$\rho_{AB}$ is steerable from Bob to Alice and from Alice to Bob.
\end{corollary}

\textit{Proof.} A two-qubit state $\rho_{AB}$ can be written in the following form under local unitary transformation,
\be
\rho_{AB}=\frac{1}{4}(\I\otimes\I+\vec{a}\cdot\vec{\sigma}\otimes \I+\I\otimes\vec{b}\cdot\vec{\sigma}+\sum \limits_{i=1}\limits^3c_i\sigma_i\otimes\sigma_i),
\ee
where $\vec{\sigma}=(\sigma_1,\sigma_2,\sigma_3)$, $\vec{a}=(a_1,a_2,a_3)$, $\vec{b}=(b_1,b_2,b_3)\in \mathbb{R}^3$ are the Bloch vectors, $a_i=\Tr[(\sigma_i\otimes\I)\rho_{AB}]$, $b_i=\Tr[(\I\otimes\sigma_i)\rho_{AB}]$, $c_i=\Tr[(\sigma_i\otimes\sigma_i)\rho_{AB}]$, $i=1,2,3$.

Let $\{P_n^{(b)}\}_{n=1}^{2}$, $b=1,2,3$, be the three MUMs with the parameter $\kappa$ constructed from the generalized Gell-Mann operators \cite{mum}, and $\bar{P_n}^{(b)}$ the conjugation of $P_n^{(b)}$. It is obvious that $\{\bar{P_n}^{(b)}\}_{n=1}^{2}$, $b=1,2,3$ are the three MUMs with the same parameter $\kappa$. We get
\begin{eqnarray*}
J(\rho_{AB})&=&\sum\limits_{b=1}\limits^{3}\sum\limits_{n=1}\limits^{2} \Tr[(P_n^{(b)}\otimes \bar{P}_n^{(b)})\rho_{AB}]\\
&=&\frac{3+(2\kappa-1)(\Tr[(\sigma_1\otimes\sigma_1-\sigma_2\otimes\sigma_2+\sigma_3\otimes\sigma_3)\rho_{AB}])}{2}\\
&=&\frac{3+(2\kappa-1)H(\rho_{AB})}{2}.
\end{eqnarray*}
According to Theorem \ref{thm:1} (Theorem \ref{thm:2}), we have $\rho_{AB}$ is steerable from Bob to Alice and from Alice to Bob, if $J(\rho_{AB})>\frac{3\mu+2\kappa-1}{2\mu}$ for $d=2$ and $\kappa_1=\kappa_2=\kappa$, which follows from $H(\rho_{AB})>\displaystyle\frac{1}{\mu}$, $\mu\in (0,\frac{1}{\sqrt{3}}]$. $\blacksquare$

In the following, we detect EPR steering of different families of two-qubit mixed states by using our results. We show by those detailed examples that our criterion based on MUMs is more convenient and operational, and more powerful than some criteria using steering inequality.

\begin{figure}
\centering
\includegraphics[width=8cm]{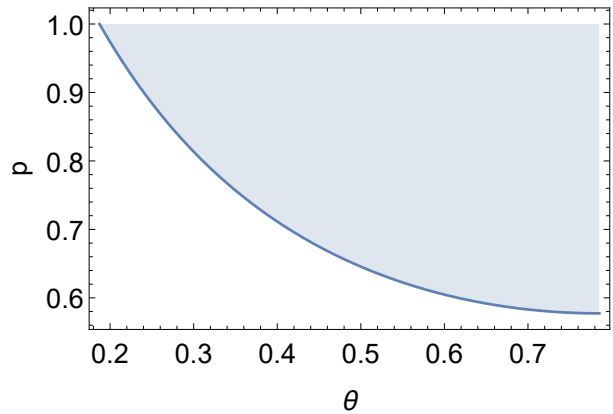}\\
\caption{\label{fig:2}The gray area represents the range of steerability that can be detected experimentally.}
\end{figure}

\textit{\textbf{Example 1.}} We consider the Werner derivative states \cite{hiro}, which are a class of non-maximally entangled mixed states and can be obtained by applying a nonlocal unitary operator on the Werner state,
\be
\rho_{\mathrm{wd}}=p|\psi_\theta\rangle\langle\psi_\theta|+(1-p)\frac{\I}{2}\otimes\frac{\I}{2},
\ee
where $|\psi_\theta\rangle=\mathrm{cos}\theta|00\rangle+\mathrm{sin}\theta|11\rangle$, $0\leqslant\theta\leqslant \displaystyle {\pi}/{4}$, $0\leqslant p\leqslant1$. From Corollary \ref{cor:1}, we have $H(\rho_{\mathrm{wd}})=p(1+2\mathrm{sin}(2\theta))$. Therefore, $\rho_{\mathrm{wd}}$ is steerable (from Alice to Bob and from Bob to Alice) for $\displaystyle {1}/{\sqrt{3}}\leqslant p<1$ and $\displaystyle \mathrm{arcsin}[\frac{1}{2}(\sqrt{3}-1)]/2<\theta \leqslant {\pi}/{4}$, see  Fig.~\ref{fig:2}. It should be noted that this steering criterion can be also derived from
the result of Ref. \cite{steer3}. However, our criteria from Theorem 1 and 2 do not need
to know the detailed state. The detection of the steerability of a state can be done by direct measurements on the state.

\textit{\textbf{Example 2.}} Consider the following class of maximally steerable mixed states (the states that violate the most the steering inequality for a given mixedness) proposed in \cite{ren},
\be
\rho^\tau=\left(
\begin{array}{cccc}
\displaystyle{(1-\tau)}/{4} & 0 & 0 &\displaystyle {(1-\tau)}/{4}  \\
 0 &\displaystyle {(1+\tau)}/{4} & \displaystyle {(1+\tau)}/{4} & 0  \\
0 & \displaystyle {(1+\tau)}/{4} & \displaystyle {(1+\tau)}/{4} & 0  \\
 \displaystyle {(1-\tau)}/{4} & 0 & 0 &\displaystyle {(1-\tau)}/{4}
\end{array}
\right),
\ee
where $-1\leqslant \tau \leqslant1$. By straightforward computation, we have that $H(\rho^{\tau})=1-2\tau>\displaystyle{1}/{\mu}$, namely, $-1\leqslant \tau \leqslant {(1-\sqrt{3})}/{2}$. Thus our criterion can detect the both-way steerability of the state $\rho_{\tau}$ for $-1\leqslant \tau \leqslant{(1-\sqrt{3})}/{2}$.
Here the upper bound $(1-\sqrt{3})/2$ is approximately $-0.366$ given in \cite{steer2}.

\textit{\textbf{Example 3.}} Consider maximally entangled mixed states presented in \cite{munro}, \be\label{munro}
\rho_{\mathrm{Munro}}=\left(
\begin{array}{cccc}
 \displaystyle h(C) & 0 & 0 & \displaystyle {C}/{2} \\
 0 & \displaystyle 1-2h(C) & 0 & 0  \\
0 & 0 & 0 & 0  \\
\displaystyle {C}/{2}& 0 & 0 & \displaystyle h(C)   \\
\end{array}
\right),
\ee
where $C$ is the concurrence \cite{con} of $\rho_{\mathrm{Munro}}$, $h(C)={1}/{3}$ for $C<{2}/{3}$ and $h(C)={C}/{2}$ for $C\geqslant {2}/{3}$.
Here, the concurrence of a pure state $|\psi\rangle$ is defined by $\displaystyle C(|\psi\rangle)=\sqrt{2(1-\Tr\rho_A^2)}$ with $\displaystyle \rho_A= \Tr_B[\rho_{AB}]$ the reduced density matrix.
The concurrence of a mixed state $\rho$ is defined by the convex roof extension: $C(\rho)=\displaystyle \mathop{\mathrm{min}}\limits_{\{p_i,|\psi_i\rangle\}}\sum\limits_ip_iC(|\psi_i\rangle)$ with $p_i>0$, $\sum\limits_i p_i=1$, and the minimization goes over all possible pure state decompositions $\rho=\sum\limits_i p_i|\psi_i\rangle\langle\psi_i|$.
It is straightforward to obtain $H(\rho_{\mathrm{Munro}})=4C-1$. Thus, the both-way EPR steerability of $\rho_{\mathrm{Munro}}$ is detected for $C>\displaystyle {(1+\sqrt{3})}/{4}\approx0.683$.

It has been shown that the state (\ref{munro}) demonstrates both-way steerability for $C > 0.707$ by using the two-setting linear steering inequality \cite{in2}. Therefore, in the region $0.683 < C\leqslant0.707$, the steerability of  state $\rho_{\mathrm{Munro}}$  can be detected by our criterion, but not by the two-setting linear steering inequality.

\subsection{Detecting  EPR Steering via general SIC-POVMs}

A POVM  $\{P_j\}$ with $d^2$ rank-$1$ operators acting on $\mathbb{C}^d$ is called symmetric informationally (SIC) complete, if
\begin{eqnarray}
P_j=\frac{1}{d}|\phi_j\rangle\langle\phi_j|,~~~~\sum_{j =1}^{d^2}P_j=\mathbb{I},
\end{eqnarray}
where $j=1, 2, \cdots, d^2$, the vectors $|\phi_j\rangle$ satisfy $|\langle\phi_j|\phi_k\rangle|^2={1}/({d+1})$, $j\neq k$.

The general SIC measurements were introduced in Refs. \cite{gsic1, gsic2}. A set of $d^2$ positive semidefinite operators $\{P_\alpha\}_{\alpha =1}^{d^2}$ on $\mathbb{C}^{d}$ is said to be a general SIC measurements if
\begin{eqnarray}
&&\sum_{\alpha =1}^{d^2}P_\alpha=\mathbb{I}, ~~~\Tr[P_\alpha^2]=a, \nonumber\\
&&\Tr[P_\alpha P_\beta]=\frac{1-da}{d(d^2-1)},
\end{eqnarray}
where $\alpha, \beta \in \{1, 2, \cdots, d^2\}$, $\alpha \neq \beta$, the parameter $a$ satisfies ${1}/{d^3} <a \leqslant {1}/{d^2}$. $a={1}/{d^2}$ if and only if all $P_\alpha$ are rank one, which gives rise to a SIC-POVM. It can be shown that $\Tr(P_\alpha)={1}/{d}$ for all $\alpha$, and general SIC-POVM can be explicitly constructed \cite{gsic2}. Let $\{F_\alpha\}_{\alpha=1}^{d^2-1}$ be a set of $d^2-1$ Hermitian, traceless operators acting on $\mathbb{C}^d$, satisfying $\Tr(F_\alpha F_\beta)=\delta_{\alpha, \beta}$. Set $F=\sum\limits_{\alpha=1}^{d^2-1}F_\alpha$. The $d^2$ operators
\be
&&P_\alpha=\frac{1}{d^2}\mathbb{I}+t[F-d(d+1)F_\alpha],~~~ \alpha=1, 2, \cdots, d^2-1,\nonumber\\
&&P_{d^2}=\frac{1}{d^2}\mathbb{I}+t(d+1)F
\ee
form a general SIC measurement. Here $t$ should be chosen such that $P_\alpha\geqslant0$ and the parameter $a$ is given by
\be
a=\frac{1}{d^3}+t^2(d-1)(d+1)^3.
\ee

Instead of the MUMs used in Theorem \ref{thm:1} (Theorem \ref{thm:2}), now we
consider the general SIC-POVMs. We have the following EPR steering criteria for qubit-qudit and qudit-qubit states. The proofs of the following theorems are similar to the case of MUMs.

\begin{theorem}\label{thm:3} Let $\rho_{AB}$ be a qudit-qubit state in $\mathbb{C}^{d}\otimes\mathbb{C}^{2}$ shared by Alice and Bob. Denote $\mathcal{P}=\{P_j\}_{j=1}^{d^2}$  and $\mathcal{Q}=\{Q_j\}_{j=1}^{4}$ two sets of general SIC-POVMs on $\mathbb{C}^{d}$ and $\mathbb{C}^{2}$ with the efficiency parameters  $a_1$ and $a_2$, respectively. Set $R_j=Q_j$ for $j=1,2,3,4$, and $R_j={\I}/{4}$ for $j=5,6,\cdots,d^2$.
Define $J(\rho)=\sum\limits_{j=1}\limits^{d^2}\Tr[(P_j\otimes R_j)\rho]$. Then, the $\rho_{AB}$ is steerable from Bob to Alice if
\be
J(\rho_{AB})>
\frac{\sqrt{\frac{a_1d^2+1}{d(d+1)}}\sqrt{\frac{4a_2+1}{6}+\frac{d^2-4}{16}}}{\mu}-\frac{1-\mu}{4\mu},
\ee
where $\mu\in (0,\frac{1}{\sqrt{3}}]$.
\end{theorem}

\begin{theorem} \label{thm:4}
Let $\rho_{AB}$ be a qubit-qudit state in $\mathbb{C}^{2}\otimes\mathbb{C}^{d}$ shared by Alice and Bob, $\mathcal{P}=\{P_j\}_{j=1}^{4}$ and $\mathcal{Q}=\{Q_j\}_{j=1}^{d^2}$ be two sets of general SIC-POVMs on $\mathbb{C}^{2}$ and $\mathbb{C}^{d}$ with the efficiency parameters $a_1$ and $a_2$, respectively. Denote $R_j=P_j$ for $j=1,2,3,4$, $R_j={\I}/{4}$ for $j=5,6,\cdots,d^2$. If
\be
J(\rho_{AB})>
\frac{\sqrt{\frac{4a_1+1}{6}+\frac{d^2-4}{16}}\sqrt{\frac{a_2d^2+1}{d(d+1)}}}{\mu}-\frac{1-\mu}{4\mu},
\ee
then $\rho_{AB}$ is steerable from  Alice to Bob,  where $\mu\in (0,\frac{1}{\sqrt{3}}]$.
\end{theorem}

As a direct application of Theorems \ref{thm:3} and \ref{thm:4}, for two-qubit states we can get the same results as the ones from corollary \ref{cor:1}.
Namely, the EPR steerable criteria based on MUMs works as well as the criteria based on general SIC-POVMs for two-qubit systems.

\section{Conclusion}\label{sec3}

We have presented criteria for detecting EPR steering of arbitrary qubit-qudit states and qudit-qubit states through MUMs and general SIC-POVMs. These criteria can be more convenient and efficient, and can be implemented experimentally. The novelty of the results is that it allows one to detect EPR steering without using the usual complicated steering inequalities. From experimental point of view, our results enable one to test EPR steering of an arbitrary qudit-qubit and qubit-qudit state directly through two classes of measurements. Our approach may be helpful to avoid the locality loophole in EPR steering test, as the degree of correlation required for entanglement testing via MUMs and general SIC-POVMs is smaller than that for violation of a steering inequality \cite{in4}. By detailed examples it has been shown that our criteria based on MUMs and general SIC-POVMs are more convenient and operational than some existing criteria in both computation and experimental implementation.

\section*{Acknowledgments}

This work is supported by the NSF of China under Grant No. 11675113, Beijing Municipal Commission of Education under Grant Nos. KM 201810011009 and KZ201810028042.

\end{document}